# Large scale quantum key distribution: challenges and solutions

QIANG ZHANG,[1,2] FEIHU XU,[1,2] YU-AO CHEN,[1,2] CHENG-ZHI PENG[1,2] AND JIAN-WEI PAN[1,2,*]

[1]*Shanghai Branch, Hefei National Laboratory for Physical Sciences at Microscale and Department of Modern Physics, University of Science and Technology of China, Shanghai, 201315, China*
[2]*CAS Center for Excellence and Synergetic Innovation Center in Quantum Information and Quantum Physics, University of Science and Technology of China, Shanghai 201315, P. R. China*
[*]*qiangzh@ustc.edu.cn*, *pan@ustc.edu.cn*

**Abstract:**
Quantum key distribution (QKD) together with one time pad encoding can provide information-theoretical security for communication. Currently, though QKD has been widely deployed in many metropolitan fiber networks, its implementation in a large scale remains experimentally challenging. This letter provides a brief review on the experimental efforts towards the goal of global QKD, including the security of practical QKD with imperfect devices, QKD metropolitan and backbone networks over optical fiber and satellite-based QKD over free space.

## 1. Introduction

Quantum key distribution (QKD) [1,2], together with one time pad (OTP) [3] method, provides a secure means of communication with information-theoretical security based on the basic principle of quantum mechanics. Light is a natural and optimal candidate for the realization of QKD due to its flying nature and its compatibility with today's fiber-based telecom network. The ultimate goal of the field is to realize a global QKD network for worldwide applications. Tremendous efforts have been put towards this goal [4–6]. Despite significant progresses over the past decades, there are still two major challenges for large-scale QKD network.

The first challenge is the gap between the theory and the practice in the actual implementations of QKD [6]. QKD is ideally secure only when it employs perfect single-photon sources and detectors. Unfortunately, ideal devices never exist in practice. As a result, device imperfections may raise security loopholes or side channels, which can break the security of practical QKD [7–13]. A solution is to design protocols that can be secure against device imperfections. Several such protocols have been proposed [14–22] and the ones that have been widely demonstrated in experiments are decoy-state QKD [19–21] and measurement-device-independent (MDI) QKD [22].

The second challenge is to go large scale, which has high channel loss and decoherence. Current distance record in fiber for QKD is 404 km [23]. At a distance of 1000 km fiber, one would detect only 0.3 photons per century, even with 10 GHz ideal single photon source and perfect single photon detectors. Very recently, a new proposed protocol is proposed to extend the distance to around 500 km [24], the key rate, however, will still drop down dramatically in long distance. One solution to this challenge is quantum repeater [25]. Although a demonstration of 500 km quantum repeater can be expected in several years [26], its real application still suffers from the limited performance of quantum memory [25]. A temporary replacement to the quantum repeater is the trustful relay scheme, which can be deployed within current technology but requires a careful protection on all the relay nodes [27]. Meanwhile, considering the much less channel loss and negligible decoherence in the space, satellite-based quantum communication is believed to be a more promising solution and has achieved lots of progresses very recently [28-30].

In this letter, we briefly review the experimental efforts over the past 30 years in solving the two challenges for global QKD. Session 2 discusses practically secure QKD with a focus on decoy-state and MDI QKD. Session 3 describes the efforts in fiber-based QKD networks that include metropolitan networks and backbone networks based on trustful relays. Session 4 reviews the very recent satellite-based QKD experiments. Session 5 provides an outlook to the global quantum communication.

## 2. Secure QKD with imperfect devices

The best-known QKD protocol is the BB84 scheme invented by Charles Bennett and Gilles Brassard in 1984 [1]. In this protocol, Alice (the sender) encodes her random numbers into a sequence of single photons prepared in different polarization states, which are chosen from two conjugate bases (rectilinear and diagonal basis), and sends the photons through the channel, fiber or free space link, to Bob (the receiver). Bob measures each incoming photon using one of the two conjugate bases. Next, Alice and Bob perform the basis reconciliation via broadcasting their basis choices via an authenticated classical channel and discard all data associated with signals prepared and measured in different bases. They sacrifice a randomly chosen portion of the remaining data to estimate the quantum bit error rate (QBER). If this quantity is larger than some prescribed threshold value, they abort the protocol. Otherwise, Alice and Bob use classical post-processing techniques (such as error correction and privacy amplification) to generate a secret key. Besides BB84, the other important QKD protocol is the entanglement based protocol proposed in 1991 by Artur Ekert [2]. Here we will focus on the BB84 protocol to discuss its security in real circumstance.

The security of BB84 is based on the quantum no-cloning theorem: an unknown quantum state cannot be perfectly copied. However, in actual implementations, the security proofs of QKD must take the theoretical models of the underlying implementation devices into consideration [15]. In general, the security proof leaves the channel to Eve [5], who can do anything over the channel as long as the physics law allows, even it is beyond the current technology. For the sender and the receiver, the security proof requires that the devices exploited in the experiment must be the same as the theoretical model. The later, however, could not be satisfied in practical circumstances, and it will leave a back door for Eve to hack practical QKD.

The first well known quantum hacking strategy is photon number splitting (PNS) attack [31,32], aiming at the imperfect photon source. The BB84 protocol requires ideal single photon source, which, however, does not exist with current technology. The current single photon source is generally bulky, expensive and low efficient. Instead, weak coherent pulses generated by highly attenuated lasers are widely exploited in QKD implementations. Since the photon number of a phase-randomized weak coherent pulse follows the Poisson distribution, there is still a probability for 2 or more photons in a pulse. Eve may exploit the multiple-photon pulses and launch the PNS attack. In this attack, Eve utilizes quantum non-demolition measurement to obtain the photon number information, blocks the 1-photon pulse and splits the pulse into two for multiple-photon pulse. Then, she keeps one part of the multiple-photon pulse and sends the other part to Bob. Later, she can get the key value during the basis-reconciliation process. In this case, Alice and Bob could not be aware of Eve's existence. The PNS attack limits the distance for QKD below 30 km [15]. Actually, in early 2000s, a few groups [33,34] have implemented QKD up to the range of 100 km with weak coherent pulses, but those systems are insecure under PNS attack.

Many QKD protocols were proposed to defend the PNS attack in early 2000. In particular, the discovery of decoy state method [19–21] made weak coherent lasers much more appealing to implement secure BB84 over long distance. In decoy state QKD, Alice prepares some decoy states in addition to the standard state – signal state – used in initial BB84. The decoy states are the same as the signal state, except for the expected photon number. Those decoy states are used for detecting Eve's attacks only, whereas the signal state is used for key generation. Each of Alice's pulses is assigned to either signal state or decoy state randomly. Alice then modulates the intensity of each pulse and sends it to Bob. After Bob acknowledges the receipt of all the signals, Alice tells Bob over an authenticated classical channel which states are signal states. The statistical characteristics can be analyzed.

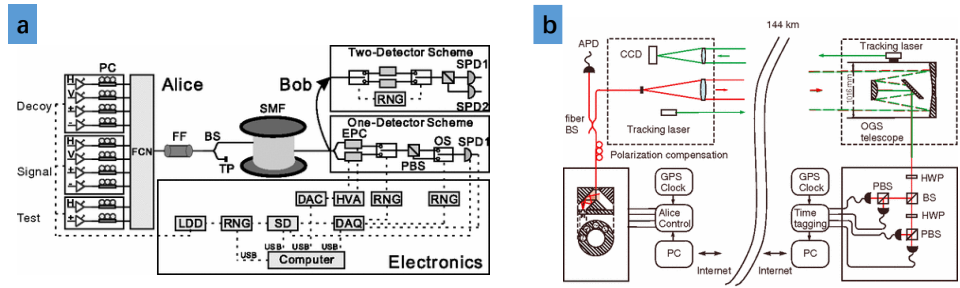

Fig. 1. (a) Decoy state QKD experiment over 100 km over fiber [36]. (b) Decoy state QKD experiment over 144 km over free space [38].

Several experimental groups have demonstrated that decoy state BB84 is secure and feasible under real-world conditions. Rosenberg et al. [35] and Peng et al. [Fig. 1(a)] [36] implement decoy state QKD through 100 km fiber, for the first time overcoming the PNS-attack-limited distance of 30 km. Note that, before the two experiments, decoy state QKD protocol has been demonstrated, which, however goes through a 15 km fiber spool [37], still less than 30 km. Later on, Schmitt-Manderbach et al. achieved 144 km decoy state QKD in free space [Fig. 1(b)] [38]. Since then, people have started to believe that QKD can be really secure with imperfect devices, and more and more experimental efforts have been made to QKD deployments in labs and field tests [39–41].

After resolving the PNS attack in the source, however, researchers started to study the loopholes in the detection side. From 2008 to 2011, several attacks against detectors have been demonstrated successfully in experiment against both research-based and commercial QKD systems. These attacks include the time-shift attack [7], detector blinding attack [9,10], dead time attack [11] and so forth [12,13]. Here, we take the detector blinding attack [9] as an example. In actual QKD implementations, the most widely used detectors by Bob are single photon avalanched diode (SPAD). When the input intensity is at single photon level, the SPAD works properly at Geiger mode. Nonetheless, when the input intensity increases, the detector will turn blind and enter linear-mode operation, i.e., the output is proportional to the input optical power. Consequently, Eve can exploit this feature to attack the QKD system. As shown in Figure 2(a), Eve can perform an intercept-and-resend attack by intercepting the pulses sent from Alice and measuring them. Then, Eve first sends a strong light to blind all Bob's detectors to linear mode, and then according to her measurement result, she sends another light at a tailored intensity to Bob such that Bob's detectors can fire only when he selects the same basis as Eve. By doing so, Eve can successfully steal the key information without being noticed.

So far, a viable solution to the detection attacks is the MDI-QKD [22]. As shown in Figure 2(b), MDI-QKD generates secret keys based on the time-reversed entanglement protocol [42,43] and leaves all the single photon detections to a public untrusted measurement platform (Eve). In MDI-QKD, both Alice and Bob are senders, and they transmit signals to an untrusted third party, Eve, who is supposed to perform a Bell state measurement (BSM). Such a measurement provides post-selected entanglement that can be verified by Alice and Bob. Since the measurement setting is only used to post-select entanglement, it can be treated as an entirely black box. Hence, MDI-QKD can remove all detection side-channels. Furthermore, MDI-QKD can be implemented with weak coherent lasers and it is practical with present technology.

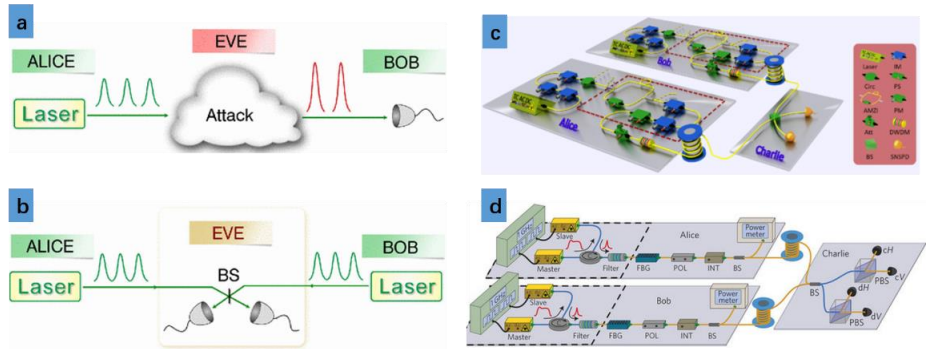

Fig. 2. (a) An illustration of the intercept-and-resend attacks against detectors (figure adopted from [44]). (b) An illustration of the MDI-QKD protocol [22]. (c) Schematic diagram of the longest MDI-QKD experiment over 404 km fiber [23]. (d) Schematic diagram of the fastest MDI-QKD experiment with 1 GHz clock rate [52].

In 2013, three experimental groups implemented MDI-QKD independently [44–46]. Liu et al. [44] achieved the first demonstration of MDI-QKD with random modulation for encoding states and decoy states. Simultaneously, Rubenok et al., demonstrated the feasibility of two-photon interference in a field test [45] and Ferreira da Silva et al., provided a two independent laser interference test in the lab [46]. Both groups also performed a partial MDI-QKD experiment, though without random modulation. Later, Tang et al., provided a full demonstration of MDI-QKD with polarization encoding qubits [47]. Wang et al., demonstrated a reference-frame-independent MDI-QKD [48] and Tang et al., demonstrated MDI-QKD with source flaws [49].

MDI-QKD is attractive not only because of its security against detection attacks, but also due to its practicality. It can resist high channel loss and reach long distance. Tang et al., implemented MDI-QKD over 200 km fiber [50] by increasing the system clock rate from 1 MHz to 75 MHz and utilizing superconducting single photon detectors. Meanwhile, Valivarthi et al., analyzed the practical issues associated with the implementation of MDI-QKD [51].

Very recently, Yin et al., extended the MDI-QKD distance to 404 km low loss fiber by optimizing the parameter and using a low-loss fiber (0.16 dB/km) [Fig. 2(c)] [23]. Importantly, the key rate achieved in the experiment at 100 km is around 3 kbps [23], which is sufficient for one-time-pad encoding of voice message. Meanwhile, Comandar et al., increased the clock rate of MDI-QKD to 1 GHz by exploiting optical seed lasers [Fig. 2(d)] [52]. Despite no random modulation on the encoding states, the 1 GHz system demonstrates the feasibility for MDI-QKD to reach 1 Mbps key rate. With all these experimental efforts, MDI-QKD is practical and suitable for metropolitan QKD network.

The above-mentioned protocols are discrete-variable QKD schemes. Another interesting option is to use continuous-variable (CV) QKD [53,54]. CV-QKD has been realized up to 100 km fiber under the security assumption of restricted (collective) attacks [55,56]. Recent theory has proved CV-QKD against general attacks [57–60], but the transmission distance in this case is highly limited [58]. On the practical security side, a major security loophole is the transmitted local oscillator from Alice to Bob, however the recent achievements on locally generated local-oscillator [61–63] provide a viable solution. Furthermore, to remove detector side channels, Pirandola et al., proposed and demonstrated a proof-of-principle CV MDI-QKD [64], but a full implementation is still a challenge in experiment.

## 3. QKD network over optical fiber: metropolitan and backbone

The immediate application of QKD is secure communication over metropolitan optical fiber network. The first reported QKD field test over optical fiber was implemented by BT Lab in Suffolk, UK [65,66], which had 4 users with one Alice and three Bobs. The quantum signal

(weak coherent pulse) sent from Alice is split into three portions by two passive beam-splitters, and then the three portions are detected by three Bobs, respectively. The length for the fiber link is around several kilometers and the secure key rate is around 1 kbps.

The first QKD network was built in 2003 supported by DARPA [67] in the city of Boston in USA. This network is a mixture of both fiber link (with optical switches) and free space link, and it is also a mixture of weak coherent state BB84 and entanglement based protocols. Later on, European fiber-based QKD network -- SECOQC -- was built in Vienna [68], and Tokyo network were reported [69]. Both networks have 6 nodes and contain multiple types of QKD protocols, including plug&play scheme, decoy state BB84, entanglement-based protocol, coherent one way (COW), differential phase shift (DPS), and continuous-variable (CV) protocol. Moreover, in both networks, beyond point-to-point link, trusted relays were used to connect remote users. In the trusted relay scenario, Alice and Bob respectively shared a secret key with a trusted relay in the middle, and then the relay will announce the XOR results of both keys publicly. With the announced result, Alice and Bob can figure out each other's key. The negative side for this method is that both Alice and Bob must trust the relay and the relay needs to be physically isolated well, while the positive side is reducing the cost and complexity as compared to the all connected point-to-point links and extending the transmission distance.

In 2009, a QKD network for real-life applications was demonstrated at the city of Hefei, China [70,71], where the voice signal of telephone was encoded and decoded by quantum keys together with the OTP method. Besides the OTP encryption, quantum keys were also combined with Advanced Encryption Standard (ASE) system, which can substantially increase the communication rate but at the cost of less security comparing to OTP. In addition, the Swiss quantum network in Geneva [72] and the Durban network in South Africa [73] reported the field test of QKD applications based on the plug&play BB84 QKD system. Furthermore, a wide area QKD network across three cities and two metropolitan areas with over 150 kilometers coverage area in China was reported [74]. Conventional QKD network architecture was reversed, where the detectors -- the most expensive components of a QKD setup -- are given to the central node and the users need only transmitters [75].

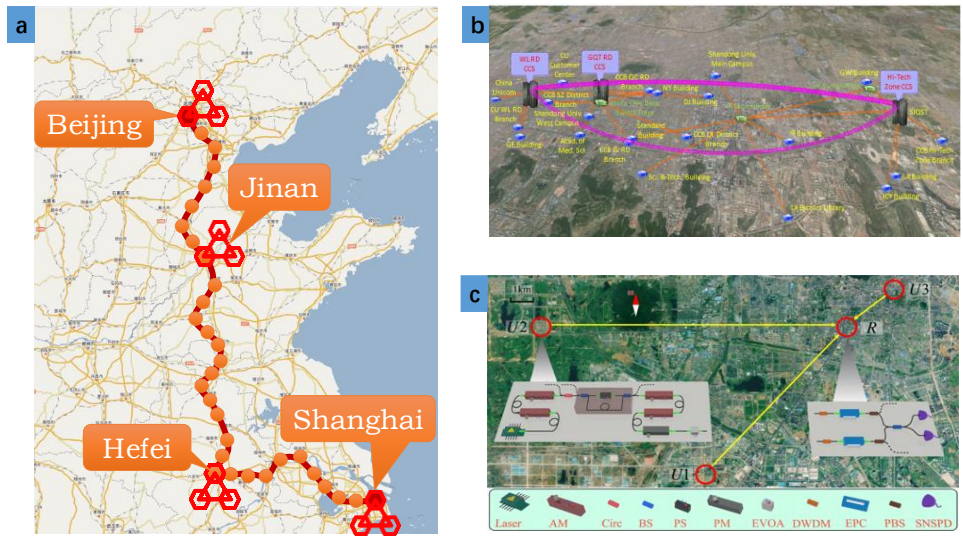

Fig. 3. (a) An outline of Beijing-Shanghai 2000 km backbone QKD network [27]. (b) The topology of Jinan metropolitan QKD network. (c) MDI-QKD network with three user nodes and one untrusted relay node [91].

Wavelength division multiplexing (WDM) is an important technique in fiber-based QKD network for optical routing [76,77] and the integration with conventional telecom network. In particular, the integration (or coexistence) between QKD and classical telecom data on a single fiber can significantly reduce the cost and increase the robustness in practical applications of QKD. The scheme of simultaneously transmitting QKD with conventional data was first introduced by Townsend in 1997 [78]. After that, several QKD experiments demonstrated the feasibility of integrating various QKD protocols with classical telecom channels [79–85]. Very recently, a QKD integrated with 3.6 Tbps traffic optical communication data over 66 km backbone field fiber has been demonstrated [86].

So far, QKD over metropolitan area has been widely deployed. Like the optical fiber telecom communication, the next step is to apply QKD to backbone optical fiber network for large-scale secure communications. The channel loss and decoherence in optical fiber, however, will reduce the intensity and fidelity of the quantum signal exponentially. Unlike the classical optical communication, the unknown quantum signal cannot be perfectly cloned (no-cloning theorem) or amplified without introducing errors. In that sense, classical optical repeater method does not work in quantum communication. As a result, loss distance QKD network is a great challenge.

Quantum repeater is thought to be one of the feasible solutions [87,88]. Although tremendous progresses have been made towards this solution, the deployment of a practical quantum repeater is still believed to be beyond current technology [89,90].

Another feasible solution in optical fiber is using trusted relays. Remarkably, China has built the world's longest quantum secure communication backbone network, from Beijing to Shanghai, with a fiber distance exceeding 2000 km [Fig. 3(a)] [27]. As shown in Figure 3(a), this backbone network includes 32 nodes in total. The backbone network connects four metropolitan networks, namely Beijing, Jinan, Hefei and Shanghai. Each metropolitan network has more than 10 nodes and different topology. As an example, the topology of Jinan metropolitan QKD network is shown in Fig. 3(b). During the operation of the Beijing-Shanghai backbone network, QKD is implemented between every two adjacent nodes and at each node, there is no quantum repeater or memory. All the 32 nodes are trusted nodes, which implement XOR operation with their received keys and pass on the quantum key along the line. The Beijing-Shanghai backbone network, together with the four metropolitan networks, is valuable for practical applications. Indeed, currently real-world applications in banks, securities, and insurances are on trial.

All the metropolitan and backbone QKD networks reviewed so far rely on trusted relays, which require that each relay node should be strictly secure. Hence, the physical isolation of the trusted nodes must be investigated, which is fatal for secure communication. A dream for future QKD network is using *untrusted* replays. In fact, the aforementioned MDI-QKD protocol provides such possibility. MDI-QKD is naturally fits a "star" type of fiber telecom network over untrusted reply, in which the untrusted Bell state measurement (BSM) can be placed on a network server in the middle and all the surrounding users just need simple transmitters that send quantum signals to the network server [6]. Any two users, who want to share secret keys via QKD, will be routed to the BSM upon their requests via an optical switch. Usually, the most expensive part of a QKD system is single photon detector in the receiver side. In this way, most of the nodes in the MDI-QKD network are senders and the entire system cost can be also reduced. Tang et al., performed the first implementation of such MDI-QKD network in the Hefei city [91], which has a star topology and four nodes with one relay node and three-user nodes [Fig. 3(c)]. The central relay node needs not to be trustful in MDI-QKD network. In practice, if the central relay can be trusted, one can reconfigure the MDI-QKD network to allow many quantum communication protocols [92,93]. Overall, among the various kinds of QKD protocols, we do believe that the MDI-QKD network will be an important choice for the future metropolitan QKD networks.

## 4. Satellite-based QKD

In general, secret key rate of QKD scales linearly with the channel transmittance, which severely limits its feasible transmission distance. As an example, QKD in standard telecomm fiber with an attenuation of ~0.2 dB/km over 1000 km results in a channel transmittance of $10^{-20}$, making the secret key rate highly low and impractical. In contrast, atmospheric attenuation in free space is less significant than in fiber. In particular, the attenuation is even negligible in the vacuum above the Earth's atmosphere whose effective vertical thickness is approximately 5-10 km from the ground. Therefore, an alternative and promising way to long distance QKD is to use satellites as intermediate trusted nodes for communication between locations on the ground -- satellite-based QKD. As long as the quantum states can survive after penetrating the Earth's atmosphere, satellite-based QKD can be far superior and more efficient than fiber links, thus offering a unique approach for quantum communication on a global scale.

In fact, the very first proof-of-concept experiment in the research field of QKD was done in free space over a link of 32 cm [Fig. 4(a)] [94]. Since then, tremendous progress that has been made in QKD experiments over free space in both laboratory and outdoor environment. Early experiments near 2000 had demonstrated the feasibility of free-space QKD over kilometer-scale distance [95–97]. In 2002, two groups independently reported the successful demonstration of QKD over distances of 10 km [Fig. 4(b)] [98] and 23.4 km [Fig. 4(c)] [99]. In 2005, Peng et al. distributed entangled photons over a two-link free-space channel to two locations separated by 10.5 km [Fig. 4(d)] [100]. Later, one-link transfer of triggered single photons over free-space channel of 7.8 km was reported [101]. These achievements proved for the first time that the quantum states, even entanglement, can still survive after penetrating the aerosphere vertically, and thus confirming the feasibility of satellite-based free-space quantum communication.

Note that the previous free-space experiments demonstrated only that QKD could survive after the absorption of the atmosphere. However, besides the absorption loss, there are also geometrical loss, optical loss, coupling loss and so forth. The total channel loss between a low Earth orbit (LEO) satellite and the Earth, or between LEO satellites, is typically around 40-45 dB [102,103]. Moreover, in previous experiments, QKD was only realized between two stationary sites, which could not simulate the real link between a fast moving satellite and the ground station.

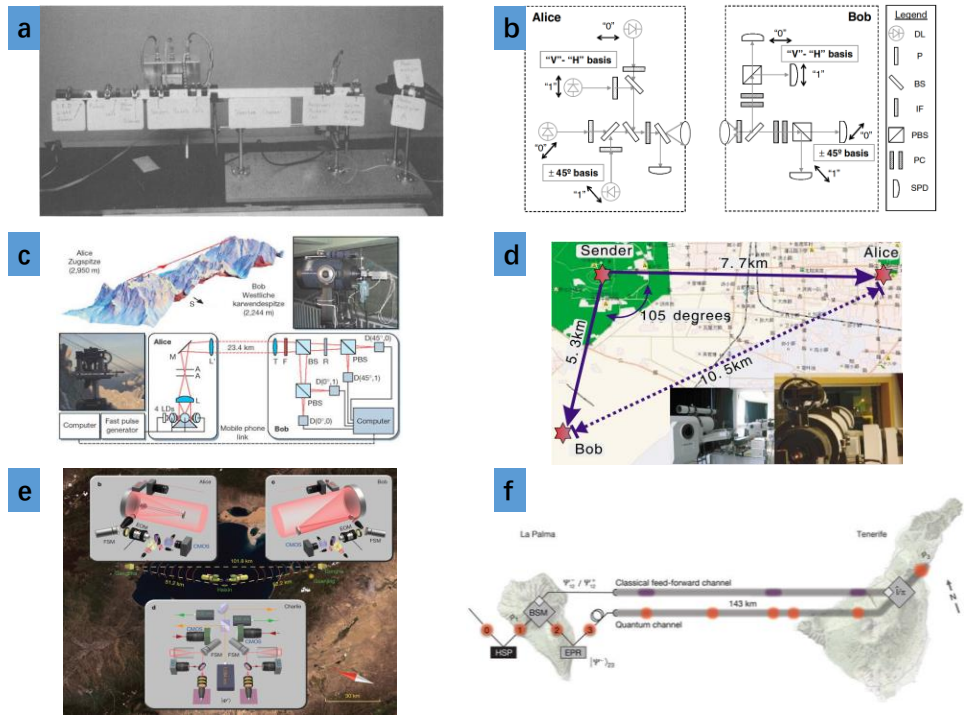

Fig. 4. (a) The first QKD demonstration over a free space link of 32 cm [94]. (b) Free-space QKD experiment over 10 km [98]. (c) Free-space QKD experiment over 23.4 km [99]. (d) Free-space distribution of entangled photons over 10.5 km [100] (e) Entanglement distribution over 100 km free space link [105]. (f) Quantum teleportation over 143 km free space link [106].

With the developments of the technologies towards the goal of satellite-based QKD realization, a series of ground tests have been performed to demonstrate the feasibility of satellite-based quantum communication before launching the real satellite. Free-space distribution of entanglement and decoy-state QKD were achieved over 144 km link in 2007 [104,38], where a secure key generation rate of 12.8 bit/s over an attenuation of about 35 dB was demonstrated in the decoy-state QKD experiment [Fig. 1(d)]. In 2012, through a two-link free-space channel with channel lengths of 51.2 km and 52.2 km respectively, entangled photons were distributed by more than 100 km [Fig. 4(e)] [105]. Moreover, two groups independently reported the teleportation of quantum states over a distance of more than 100 km free-space link [Fig. 4(f)] [105,106].

Direct and full-scale experimental verifications have been performed towards the goal of satellite-based QKD for moving platforms [107]. In this experiment, the system was operated in a moving platform through a turntable and floating platform of hot-air balloon over a huge lossy channel, for substantiating performances under rapid motion, altitude change and vibration, random movement of satellites and in high-loss regime, which covers almost all the leading parameters of a typical LEO satellite. Meanwhile, Nauerth et al. reported the remarkable demonstration of QKD between an airplane and a ground station [108]. Later, a direct demonstration of the satellite-ground transmission of a quasi-single-photon source, which is generated by reflecting weak laser pulses back to earth with a cube-corner retro-reflector on the existing satellite, was reported [109]. In 2015, quantum communication from space to ground via the in-orbit satellite corner cube retroreflectors was demonstrated by G. Vallone et al. [110].

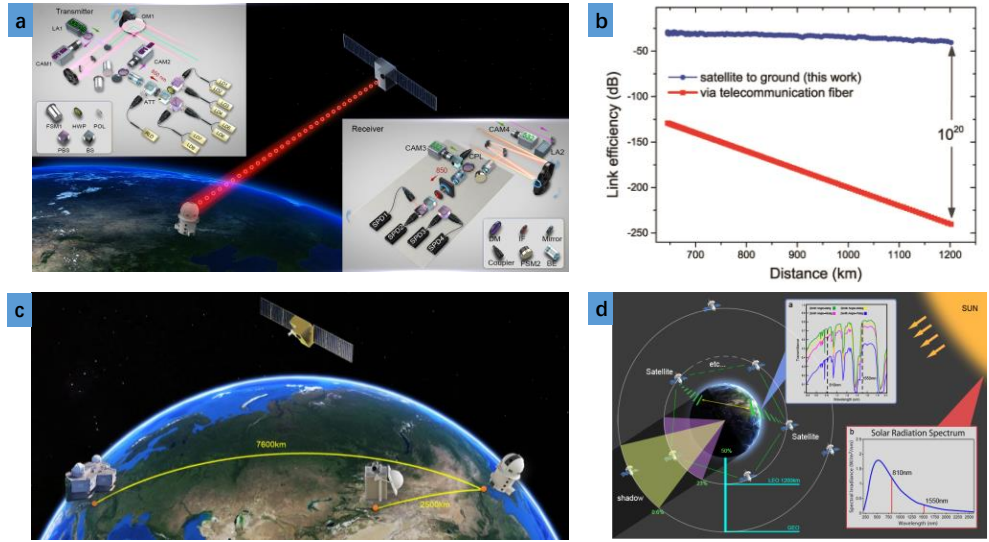

Fig. 5. (a) An overview of the satellite-to-ground QKD [28]. The decoy BB84 transmitter is on the satellite to perform downlink QKD and the receiver is located at Xinglong observatory station. (b) Link efficiencies comparison between direct transmission through telecommunication optical fibers (red) and the satellite-to-ground approach (blue) [28]. (c) Intercontinental QKD between China and Europe [111]. (d) An illustration of global QKD with LEO satellites and geosynchronous equatorial orbit (GEO) satellites [112].

In August 2016, the first quantum science satellite -- Micius -- was launched in Jiuquan, China, which is a LEO satellite orbiting at an altitude of about 500 km. As is shown in Fig. 5(a), a decoy-state QKD transmitter is equipped on one of the Micius satellite payload, and it performs downlink QKD by sending the quantum signals to the receiver located at Xinglong ground station (40 °23045.1200N, 117 °34038.8500E). The decoy-state QKD was demonstrated with polarization encoding from the satellite to the ground, with 1 kbps rate over a distance of up to 1200 km [28]. The QKD experiment was performed on 23 different days, and the quantum bit error rates and secure key rates varied with the distance and weather in different days. Fig. 5(b) compared the optical link efficiency over distances ranging from 645 km to 1200 km between the satellite-based QKD and the conventional method of direct transmission through optical telecommunication fibers. Despite the short coverage time (273 s per day) and the need for reasonably good weather conditions using the Micius satellite, a substantial enhancement in efficiency compared to telecommunication fibers was observed. At 1200 km, the channel efficiency of the satellite-based QKD over the 273-s coverage time is about 20 orders of magnitudes higher than previously reported QKD experiments using optical fiber. Meanwhile, based on the Micius quantum satellite, the ground to satellite quantum teleportation [29] and the satellite-based entanglement distribution over 1200 km were also demonstrated [30].

Later, the Micius quantum satellite was used as a trusted relay to distribute secure keys between multiple remote locations in China and Europe [Fig. 5(c)] [111]. The three cooperating ground stations were located at Xinglong, Nanshan (43 °28031.6600N, 87 °10036.0700E), and Graz (47 °401.7200N, 15 °29035.9200E). The distances from Xinglong to Nanshan and to Graz were 2500 km and 7600 km, respectively. In this experiment, Micius flied along a Sun-synchronized orbit, which circled Earth every 94 min. Each night starting at around 0:50 a.m. local time, Micius passed over the three ground stations allowing for downlink QKD in a duration of ∼300 s. Under reasonably good weather conditions, sifted key rates of ∼3 kbps at ∼1000 km physical separation distance and ∼9 kbps at ∼600 km distance (at the maximal elevation angle) could be routinely obtained. Therefore, using the Micius quantum satellite as a trusted relay, the first intercontinental quantum communication was established successfully.

## 5. Outlook of the future global QKD

In summary, the decoy state QKD and the MDI-QKD provide viable and practical solutions for imperfect source and detectors to realize practically secure QKD in experiment. A simple prototype for a global quantum communication network based on satellite and backbone fiber network has been demonstrated. In the future, more and more backbone and metropolitan fiber QKD networks are expected to be built in China and Europe for applications. To increase the coverage time and area for a more efficient satellite-based QKD network, one can launch higher-orbit quantum satellites and implement QKD in daytime using telecommunication wavelength photons and tighter spatial and spectral filtering [Fig. 5(d)] [112]. One limitation of the current implementation of the global QKD is that the satellite itself should be trusted. This can be overcome in the future using entanglement-based QKD systems [30,113] or MDI-QKD fiber networks. In the long run, device independent QKD [16,17] could be implemented based on the loophole-free Bell test where the users do not even need to trust their implementation devices, though it seems still not practical with current technology.

**Funding**

This work has been supported by the National Key R&D Program of China (2017YFA0307900); the Strategic Priority Research Program on Space Science; Chinese Academy of Sciences; the National Natural Science Foundation of China.

**Acknowledgment**

The authors thank Yang Liu, Wen-Zhao Liu, Wen-Qi Cai, Sheng-Kai Liao and Juan Yin for their valuable comments and suggestions.